# *In aqua* electrochemistry probed by XPEEM: experimental setup, examples, and challenges

Slavomír Nemšák,[1,†] Evgheni Strelcov,[2,3] Hongxuan Guo,[2,3] Brian D. Hoskins,[2] Tomáš Duchoň,[1,4] David N. Mueller,[1] Alexander Yulaev,[2,3] Ivan Vlassiouk,[5] Alexander Tselev,[6] Claus M. Schneider,[1] and Andrei Kolmakov[2,‡]

[1]Peter-Grünberg-Institut 6, Forschungszentrum Jülich GmbH, 52425 Jülich, Germany

[2]Center for Nanoscale Science and Technology, NIST, Gaithersburg, MD 20899, USA

[3]Maryland NanoCenter, University of Maryland, College Park, MD 20742, USA

[4]Department of Surface and Plasma Science, Faculty of Mathematics and Physics, Charles University, Prague 18000, Czech Republic

[5]Oak Ridge National Laboratory, Oak Ridge, TN 37831, USA

[6]CICECO-Aveiro Institute of Materials and Department of Physics, University of Aveiro, 3810-193 Aveiro, Portugal

## Abstract

**Recent developments in environmental and liquid cells equipped with electron transparent graphene windows have enabled traditional surface science spectromicroscopy tools, such as X-ray photoelectron spectroscopy (XPS), photoemission electron microscopy (PEEM), and scanning electron microscopy (SEM) to be applied to study solid-liquid and liquid-gas interfaces. Here, we focus on the experimental implementation of PEEM to probe electrified graphene-liquid interfaces using electrolyte-filled microchannel arrays as a new sample platform. We demonstrate the important methodological advantage of these multi-sample arrays: they enable the combination of the wide field of view hyperspectral imaging capabilities from PEEM with the use of powerful data mining algorithms to reveal spectroscopic and temporal behaviors at the level of the individual microsample or the entire array ensemble.**

## 1. Introduction

Understanding near-electrode properties, such as ion densities, electric potential distribution within double layer, specific or non-specific ion adsorption, and redox reactions at the solid-liquid interfaces, is a subject of active fundamental and applied studies. Since the major electrochemical (EC) processes are interfacial in nature and take place within a few nanometer-thick layer near the electrode, the crucial requirements for obtaining a detailed picture of the interface are: (i) an adequate information depth (the maximum depth from which the spectra or images can be recorded) to access the buried layers and (ii) a sufficient spectral and depth selectivity to be able to analyze them. In addition, the kinetics of physicochemical reactions at the interfaces often requires *in operando* and simultaneous multi-parametric (e.g. spectral and potentiometric) measurements. Finally, the spatial inhomogeneity of the interfacial phenomena, such as the nucleation and growth of solid products, the distribution of defects and adsorption sites at micro- and nanoscales, necessitates the application of microscopic techniques.

A number of experimental approaches has been developed in the past few decades to probe solid-liquid interfaces and liquid electrolytes under operating conditions.[1, 2] However, many of these techniques fall short of meeting the aforementioned requirements. For example, X-ray absorption spectroscopy, which relies on the fluorescent yield as a measure (as well as many of the photon-in/photon-out methods), has a sufficient information depth (from a few nanometers to several micrometers), but performs poorly in terms of the depth selectivity. There are modifications of these techniques which improve the depth sensitivity and selectivity, such as total electron/ion yield collection[3] or X-ray standing wave approach in both the soft[4] and tender X-ray regimes.[5] These approaches can overcome this problem, but require specially prepared samples and are not universal.

On the other hand, analytical approaches, such as Ambient Pressure X-ray Photoelectron Spectroscopy (APXPS) based on characteristic photoelectron detection, can yield an exceptional depth/surface selectivity (down to ≈0.1 nm regime) at interfaces. This is due to both the short electron inelastic mean free path and the exponential dependence of photoelectron intensity on the probing depth in a liquid sample.[6-9] An important addition possessing all of the benefits of the X-ray absorption and photoemission spectroscopy is the capability of (X-ray) photoelectron emission microscopy ((X-)PEEM) to image surfaces and buried interface with high spatial (nanoscale) and temporal (femtosecond) resolution.[10] By combining micro-focus beamlines and tunable light sources, this technique can provide unique information on the chemical composition, morphology, electric field profile, work function, ferroelectric, ferromagnetic and antiferromagnetic properties, *etc.* – a broad characterization, which is very hard to obtain with any other individual technique (see a recent review[11] and references therein). However, the application of this technique to solid-gas and even more challenging solid-liquid interfaces is hampered by several experimental limitations. The most important of them is the fact that the sample serves as a cathode for the immersive optics, and as such, it is at high negative potential with respect to the detector. Extractor fields up to $10^7$ V/m are used, and the probability of an avalanche electrical discharge is critically high therefore over a wide range of pressures. One solution, differential pumping, has successfully been realized in PEEM with pressures up to $10^{-1}$ Pa,[12] but has not been able to operate at the higher pressure range (≈ $10^3$ Pa) routinely achievable using APXPS. In addition to early in-transmission PEEM designs [13], ongoing development of the next generation of elevated pressure, commercial, differentially pumped photoelectron emission microscopes deserve mention as well

Recent advances in microfabrication of ultrathin free-standing silicon nitride ($Si_xN_y$), silicon oxide ($SiO_2$) and silicon carbide (SiC) membranes, and especially the advent of two-dimensional (2D) materials such as graphene, have paved the way towards a resolution of the pressure gap problem in ambient pressure PEEM studies. The electron transparency, molecular impermeability, and mechanical strength of graphene allows the study of intercalated and graphene encapsulated samples[14] including ultrahigh pressure gases.[15] More recently, a universal sample platform to study arbitrary liquid samples with PEEM has been proposed.[16] This platform is based on microfabricated microchannel arrays filled with liquids of interest and capped with bilayer graphene to isolate the liquid content from the ultra-high vacuum (UHV) environment. In this report, we describe the application of this approach to study electrified graphene-liquid interfaces.

## 2. Through-the-membrane XPEEM: general considerations

There are numerous examples of PEEM imaging of buried interfaces through oxide layers and thin films in a wide photon energy range spanning from ultra-violet (UV) to hard X-rays.[17-19] Probing a liquid solution through an ultrathin membrane is not principally different, although it imposes some thickness, strength, and material restrictions on such a membrane. As mentioned above, the relatively small inelastic mean free path (IMFP) of the X-ray generated photoelectrons in condensed matter determines the depth sensitivity for all electron photoemission techniques. The dependence of IMFPs on the kinetic energy of photoelectrons in carbon/graphite, silicon and liquid water is shown in Figure 1a,b. These substances are chosen because they are the most common materials currently used for membrane fabrication in (fluidic) liquid cells, as well as in-liquid transmission electron microscopy. In addition, graphene-based materials were used as a capping membrane in Refs.[16, 20-23], as an electron transparent working electrode in liquid cells,[24, 25] and also in another example shown later in this paper.[26].

Consider a 1 nm thick interfacial double layer (DL) region in a liquid adjacent to a solid Si membrane, through which the emitted photoelectrons are collected (see the scheme in Figure 1a). This experimental geometry facilitates a normal electron emission and a shallow X-ray incident angle typical of synchrotron photoemission facilities. The O 1s photoelectrons with a binding energy of approximately 530 eV are selected as a reference core-level since they are widely used for analyzing aqueous solutions. Their chemical shifts provide well distinguishable states for water vapor, liquid, and other oxygen containing species, such as oxides and hydroxides. Figure 1b shows that the expected photoemission information depth strongly depends on the energy of the photons and practically feasible thickness of the membranes roughly scales as 1 nm, 5 nm and 10 nm for the soft, tender and hard X-rays, respectively. Commercially available Si, SiC, $SiO_2$ or $Si_xN_y$ membranes with thicknesses of ≈10 nm are acceptable for tender and hard X-ray photoemission studies with liquid cells[27], but much thinner, one- or a few-layer graphene membranes, are needed to perform experiments with soft X-rays.

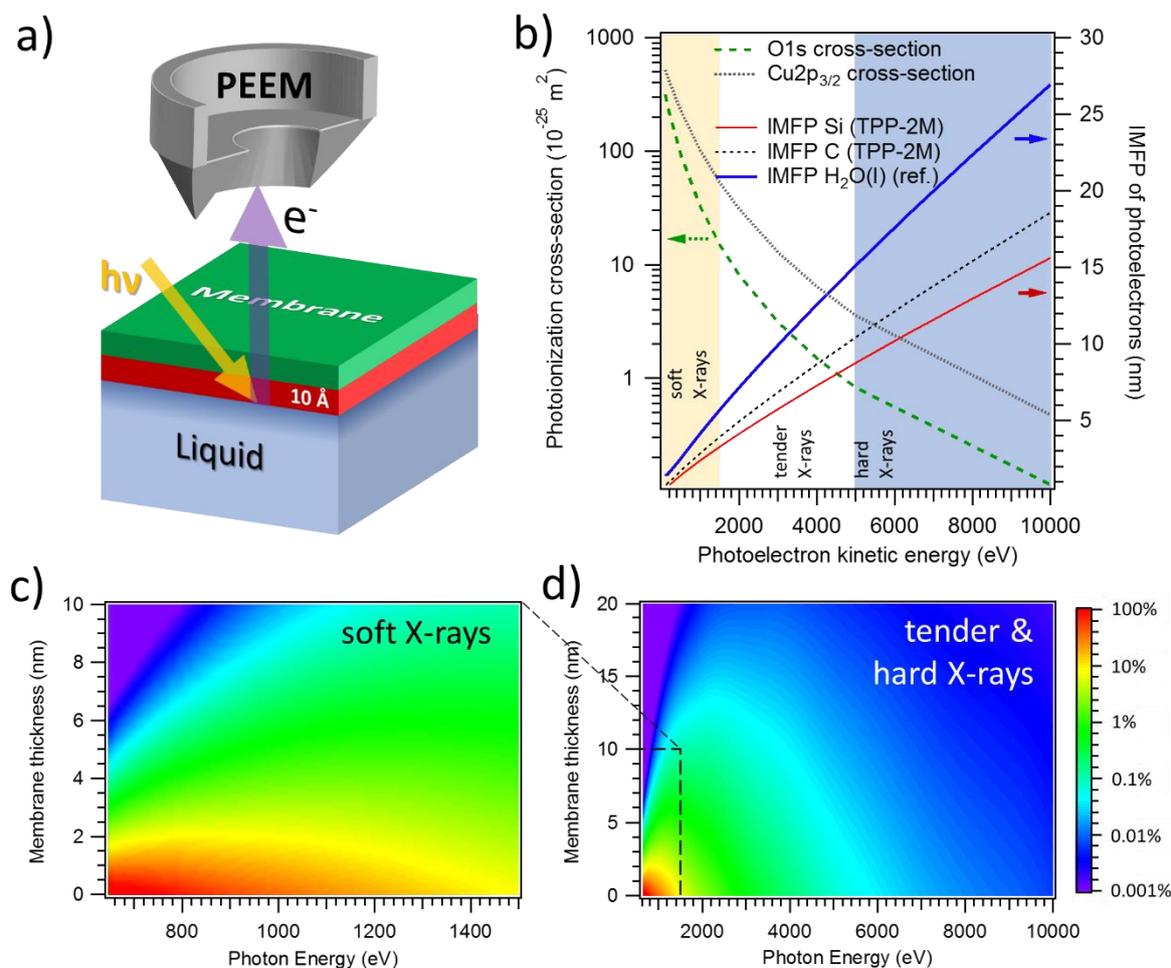

**Figure 1.** a) Experimental setup for through-the-membrane PEEM in liquids; (b) Left axis: energy dependence of the photoionization cross sections for O 1s and Cu 2p3/2 levels. Right axis: energy dependence of the IMFP values for graphite and silicon and water. The values for C and Si were obtained using relativistic TPP-2M formula[28], values for water are taken from Ref. [29]. (c, d) Simulated water O 1s photoelectron yield (in percent of max O 1s intensity) as a function of photon energy and membrane thickness for soft (c) and tender/hard (d) X-rays (see details in the text).

Apparently, the higher the photon energy the larger the information depth can be achieved. However, the fast drop of the photoionization cross section can negate this effect and a trade-off between the optimum energy range, materials to study, and the membrane thickness must be found. As an example, Figures 1c,d demonstrate the dependence of the O 1s photoemission signal originating from the aforementioned thin interfacial liquid layer on the excitation energy and the solid membrane thickness. The simulation includes an assumption of the exponential attenuation of the photoelectron signal and uses the information on the IMFPs and photoionization cross-sections provided in Figure 1b. The soft X-ray region of energies, depicted in Figure 1c, reveals that the maximum usable thickness of a Si membrane for the lowest displayed energies is between 1 nm and 1.5 nm. A thicker membrane results in O 1s signal intensity attenuation by more than a factor of 10 compared to the signal from the uncovered liquid layer. With increasing excitation energy, the maximum usable membrane thickness increases as well and the optimum energy range for probing through a commercially available ≈10 nm thick membrane lies between 2000 eV and 4000 eV. Using these excitation energies still brings a decent interfacial sensitivity, while sufficiently increasing the information depth. A further increase of the photon energy extends the information depth even further, but both a rapid decrease of the photoionization cross-sections and the fact that the experiment becomes more bulk-sensitive causes a drastic drop of the overall signal from the interfacial region.

### 3. EC-cell design considerations

The only known conducting materials satisfying the membrane thickness requirement for soft X-rays photoemission are mono- or bilayer graphene and possibly carbon nanomembranes.[19] The mechanical stability of such a thin membrane defines the lateral size of the electron transparent window. In the case of a single-crystal monolayer graphene, a window with a few micrometers in diameter is able to sustain a pressure difference of several bars.[30] Chemical vapor deposition (CVD) grown graphene[2] and the aforementioned carbon membranes[3] provide somewhat reduced performance. Therefore, device geometries for a liquid cell are restricted to a single or multiple micro-orifice front surface covered with graphene. Multiple micro-orifice array-like E-cells (Fig. 2) appear to be more practical since the disruption of one or few graphene windows during the fabrication process or experiment will not terminate the entire array. However, the high vacuum requirements of PEEM impose severe leakage restrictions on the design of such an array. In fact, assuming a molecular gas flow $Q = 0.25 \cdot \Delta P \cdot S \cdot \sqrt{kT/m}$ from the leaking (broken) orifices (here $\Delta P \approx 2000$ Pa is the pressure difference and $S$ is the area of the opening) having the pumping speed $R \approx 0.2$ m$^3$/s one can maintain the vacuum pressure $P = Q/R \approx 10^{-5}$ Pa only if the total area of broken orifices $S$ will be below 10 µm$^2$. This area corresponds to only two-three orifices with 2 µm in diameter.

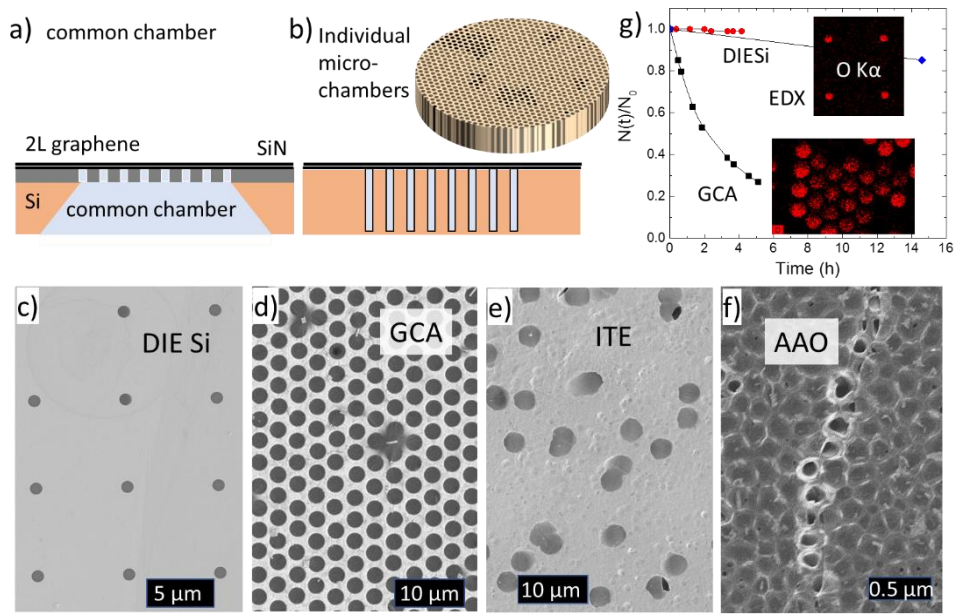

**Figure 2**. Common chamber (a) and microchamber array (b) designs of the graphene liquid cells; (c-f) Different graphene-capped micro porous matrixes which can be used as multichannel arrays: deep ion etched Si wafers (c), glass capillary microarrays (d), ion track etched polymer membranes (e), porous anodic alumina membranes (f). (g) Time dependence of the filling factor (the ratio of the channels retaining water at time t: (N(t)) to initial number of water filled channels (N$_0$) for two different MCA designs. Insets depict corresponding X-ray maps recorded at O Kα.

There are two major designs of the multi-orifice graphene liquid cells (Fig. 2a). In the E-cell design, where all graphene capped orifices connect to a common flow chamber (so-called common chamber design), the disruption of even a few of them raises the steady state pressure in the chamber to $10^{-5}$ $10^{-1}$ Pa making PEEM studies impossible in its standard configuration, and requiring the system to operate under differential pumping conditions.[21, 24] As an alternative for standard XPS and PEEM systems, we implemented a UHV compatible E-cell design based on a planar array of separated microchambers filled with electrolyte and capped with bilayer graphene (Fig. 2b).[31] The advantage of this sample platform is that an accidental or beam-induced disruption of the graphene results in only a miniscule amount of liquid (the volume of the individual microchamber) being sprayed into the vacuum chamber. Consequently, UHV conditions inside the PEEM chamber can be preserved. There are many nano/micro porous matrices which can be graphene-capped and used for these applications. These include deep ion etched Si wafers, glass capillary microarrays, ion track etched polymer membranes, porous anodic alumina membranes, and other possible systems (Fig. 2 c-f). The lifetime of the liquid inside this sample platform depends on the total volume of liquid contained inside the individual microchamber and the liquid runaway rate due to both parasitic diffusion at the graphene-support interface and through graphene defects. The leakage can be significantly reduced by using bilayer graphene instead of a single layer one, improving the graphene-substrate adhesion and increasing the interfacial diffusion paths. For the tested microchannel array (MCA) designs depicted in Figures 2c, d with individual microchamber volumes of $\approx 4 \cdot 10^3$ µm$^3$ and $4 \cdot 10^4$ µm$^3$, respectively, the measured lifetime of liquids varies between a few and tens of hours, which is sufficient for a routine PEEM experiment (Fig. 2g). The buildup of the radiolitic products inside the channels and at the graphene-support interface is among the factors limiting the performance of the MCA platform, contributing in the long term to bubble formation and graphene delamination.

The fabrication of the MCA E-cell proceeds through the following major stages (Fig. 3a). First, the bottom counter (Pt) and top (Au/Cr) electrodes are deposited via atomic layer deposition (ALD) and sputtering processes, respectively. It is essential to metalize at least one third of the interior length of the channel from the bottom to ensure a reliable contact between the slowly evaporating electrolyte and counter electrode. Bilayer graphene is then transferred onto the Au coated MCA surface using a standard PMMA based protocol. After annealing the graphene-PMMA stack, the PMMA is dissolved in a large amount of acetone. The acetone in the channels is subsequently substituted with water without removing the sample from the liquid environment. This drastically improves the yield of the liquid filled cells. The electrolyte of the desired composition and concentration is then drop-casted onto the back of the MCA. The volume of the electrolyte necessary exceeds at least by two orders of magnitude the volume of water contained inside MCA. After concentration equilibrium is reached, the excess of the electrolyte is removed using filter paper and the cell is sealed using a water immiscible UV curable adhesive.

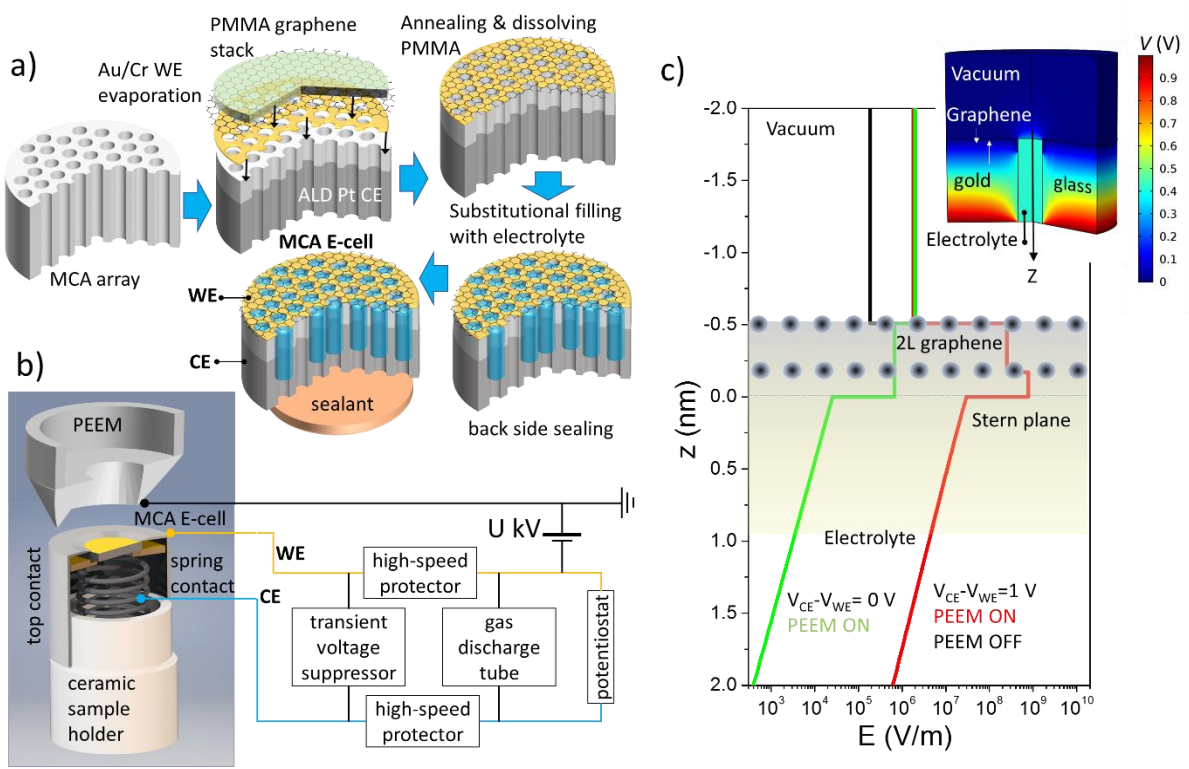

**Figure 3.** a) MCA E-cell fabrication steps (see details in the text), CE and WE stand for working and counter electrodes, respectively; b) Electric setup for electrochemical PEEM; c) Finite elements modelling of the electric field strength across the bi-layer graphene-electrolyte ($10^2$ mol/m$^3$ CuSO$_4$) interface under electrochemical PEEM settings. Red and green curves correspond to "PEEM ON" conditions when high $2 \cdot 10^6$ V/m field exists between immersion lens and graphene working electrode. Black curve corresponds to conditions when "PEEM is OFF" but 1 V potential difference between working WE and counter CE electrodes retains. The gradient shadow area corresponds to XPEEM probing depth. Inset is a finite elements model layout for MCA E-cell polarized at 1 V.

The E-cell sample holder design is shown in Figure 3b. An MCA E-cell chip carrier, which is a double-sided printed circuit board with suitable electrical connections providing a contact to the bottom counter (CE) electrode, serves also as an axis centering piece and a support of the multi-orifice array electrochemical cell. If needed, the E-cell and the chip carrier can be mechanically bonded using silver paint or other UHV compatible bonding means. The chip carrier can also accommodate uneven bottom surfaces of the electrochemical cells which often occur due to the bottom seal. In that case, the central part of the chip carrier can be carved out using a milling machine. Additionally, the developed chip carriers already have separate electrical contacts for a prospective 3-electrode cell design. The top working electrode (WE) of the cell is in direct contact with the metallic cover, which also closes the top part of the sample holder and acts as a counter-piece for the spring-loaded mechanism.

The electrical connections and sample biasing in photoemission microscopes have additional challenges which are not present in other UHV spectro- and microscopic techniques. As discussed earlier, the sample in PEEM is a part of the immersion lens (IL) and it is electrically connected to a high extracting voltage or to a high potential determined by the kinetic energy of the energy filter. A power supply (or a potentiostat) used for sample biasing therefore must be floating under this high potential. This creates complications in the power supply control, in the electrical output, and its protection (Fig. 3b).

The remote control and read-out of such power supplies is usually done via an optically decoupled connection or via some standard wireless connection. On the other hand, the output stages of these power supplies are usually equipped with surge protections. This is important especially in the case of a liquid containing cell, since an abrupt failure and the loss of the structural integrity of the sealing membrane creates a pressure burst in the vacuum chamber and consequently precipitates an electrical discharge between the sample and the extractor lens. The surge protection needs to be fast enough to protect the output stage of the amplifier and at the same time it should not create parasitic currents and increase noise levels significantly.

The low-energy electron microscopy (LEEM)/PEEM endstation of BESSY-II, at which the described experiments took place, is equipped with the commercial potentiostat specially customized for this purpose. The potentiostat is encased in a high voltage rack and one of its outputs is electrically connected to the high negative potential of the sample. The same contact is then connected to the top graphene electrode of the electrochemical cell. Another output contact is electrically connected to the bottom platinum electrode (Fig. 3b). Communication is carried over the fiber Ethernet using fiber/copper media converters. The electrical output stages use a cascade surge protection that combines gas discharge protectors, high-speed metal–oxide–semiconductor field-effect transistor (MOSFET) semiconductor over-current/overvoltage protectors, and transient voltage suppression diodes. All electronic parts were selected with regards to the low capacitance and low leakage currents of the output stage. A block diagram of the surge protections for two output channels is shown in Figure 3b. Ground potential in the surge protection unit is connected to a high voltage isolated ground. The potentiostat is able to produce output voltages in the range of +/- 10 V and currents up to 100 mA with an accuracy down to 10 pA.

Another consequence of the sample being a part of the immersion lens in the PEEM setup is the presence of an appreciable electric field with a strength of few megavolts per meter at the surface of the graphene. It is important to determine whether such a strong field can noticeably perturb the state of the electrolyte behind the graphene. To estimate the degree of the influence of few kV potential difference between the IL and the graphene surface, we performed numerical simulation of the electric field distribution across the liquid cell-PEEM system using finite elements (FE) modeling. The graphene membrane was assumed to be made of two uncoupled and undoped single-layer graphene layers separated by a distance of 0.34 nm. The graphene relative dielectric permittivity was set to 4.[32]

To calculate the screening of the electric charge and field by graphene, we used the model developed in Ref.[33] assuming that the smearing of the step-like Fermi distribution at room temperature is small compared to the Fermi level shifts associated with the experimental potential variations greater than 150 mV. The Gouy-Chapman theory was employed to describe the 2:2 electrolyte. The FE model layout is displayed in the inset of Fig. 3c. To represent the electric field created by the potential difference between the IL and the graphene surface, a suitable electric potential was set as a boundary condition on the top surface of the model.

The calculated electric field distribution along the channel axis in the vicinity of the graphene membrane are shown in Fig. 3c. As can be seen, the IL-graphene potential (and the field E ~ $2 \cdot 10^6$ V/m on the top surface of the graphene membrane) has negligible effect on the state of the DL in the electrolyte, when a bias voltage of 1 V applied to the bottom electrode of the liquid cell. The plots in Fig. 3c show that the field at the electrolyte-graphene interface is about 100 times larger than at the top graphene surface, which makes the effect of the immersion lens field negligible. When the bias voltage is zero, the external electric field induces ion redistribution in the electrolyte due to incomplete screening of the field by graphene. However, the strength of the electric field in the DL beneath the graphene is about three orders of magnitude smaller than in the case when 1 V is applied to the bottom electrode.

## 4. The importance of the MCA platform for high throughput PEEM data acquisition and mining

Unlike environmental cells used in electron microscopy or X- ray spectroscopy that have one common chamber for a single or multiple windows, the MCA platform offers a group of thousands of independent liquid-filled microchannels with minimal cross-talk between the neighbors. Thus, in this configuration, the experiment is performed not just on one sample, but also on an ensemble of samples. In addition, various regions of one MCA chip can be filled with different electrolytes, or a gradient of a property of interest (e.g. concentration) can be created across multiple channels in a small region of the MCA and combinatorically monitored in the wide field of view of PEEM. These advantages, however, come with a challenge: the large hyperspectral datasets require tools for being visualized, processed and interpreted.[34] The natural choice of such tools is the multivariate statistical analysis toolbox including such methods as principle component analysis (PCA), independent component analysis, clustering algorithms, and Bayesian inference methods. These techniques can help reduce the data dimensionality, denoise the data, visualize a multidimensional dataset, unearth a statistically significant behavior or trend, evaluate the data quality, and estimate the measurement uncertainty. An important aspect of data analysis is its interpretation, which is commonly done via fitting the experimental data to a postulated physical model. However, this process introduces a subjective bias into the original data, and oftentimes minor discrepancies between the data and the model fit are ignored or go unnoticed even if they contain important information about the studied system. The multivariate methods, on the other hand, are non-discriminatory, highlighting every statistically significant trait in the data and giving the researcher the opportunity to visualize it and contemplate it. The introduction of physical constraints (of a much more general nature than the specific physical models) into the sophisticated statistical methods may also allow for finding a clear physical meaning to these traits. Finally, large statistical datasets can form combinatorial libraries of behaviors which can be used for non-classical unbiased modeling, resampling, and forecasting utilizing the neural network approaches. Although the full realization of the described strategies is still in the future, the developed PEEM MCA setup allows for making the first large step in this direction: the collection of a statistical ensemble of hyperspectral data containing information on the electrochemical behavior of the studied system. Below we will show how to use Bayesian Linear Unmixing (BLU) to denoise and visualize such datasets. The BLU algorithm splits a 3D dataset into a linear combination of a user-defined number of position-independent spectral components (endmembers - S) and corresponding abundance maps (A), simultaneously filtering out noise (N) such that: *I(x,y,t) = S(t)·A(x,y)+N*. A generic BLU analysis flowchart is depicted in Figure 4 and the details of this method can be found elsewhere.[35-37]

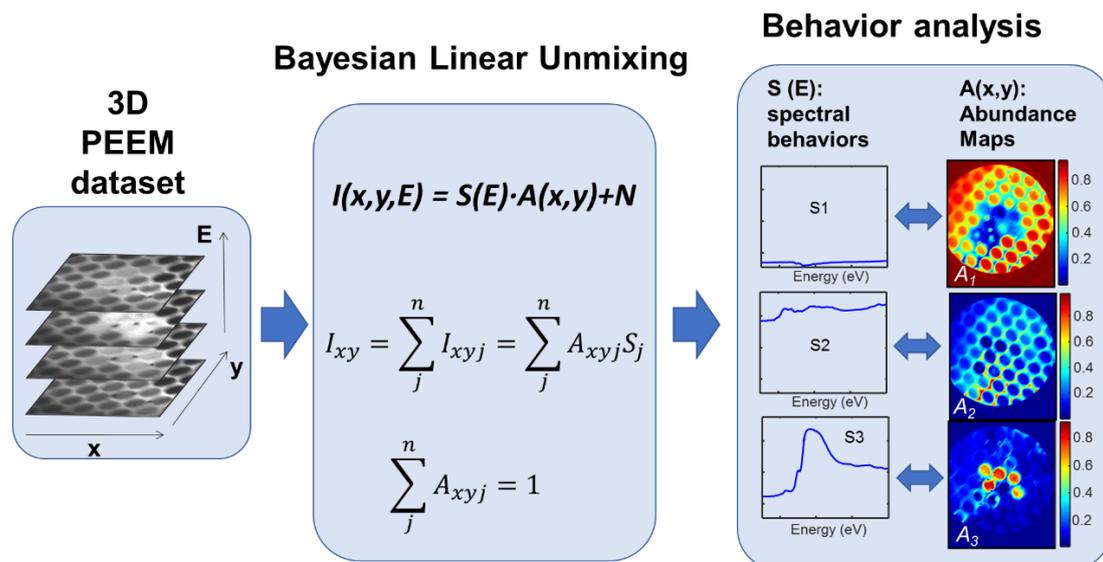

**Figure 4.** Bayesian Linear Unmixing of PEEM dataset: A 3-dimensional dataset is split into a linear combination of spectral endmembers (*S*) and abundance maps (*A*) with simultaneous noise (*N*) filtering. The spectral behaviors and their spatial distributions can be then analyzed to properly assign materials/system components that manifest such behaviors. In the shown case, $S_1$ is the spectrum of empty channels and microscope aperture, $S_2$ is behavior of the gold-coated MCA frame, and $S_3$ is the spectrum originated in the water-filled channels.[31]

It is noteworthy, that due to the large number of individual cells in the MCA platform, a gradient of different half-cell potentials can be applied and investigated simultaneously, in line with the discussed combinatorial approach. While combinatorial approaches are not new to the field of electrochemistry, previous investigations have mostly focused on continuous variation of material chemistry[38, 39] and analysis of the resulting electrochemistry, or the electrochemical preparation of graduated materials by elaborate geometries.[40] All of these concepts have a common feature of decoupled (electrochemical) preparation and (electrochemical) analysis, which requires very well defined geometries in order to allow the identification of unambiguous structure-property relations. Our approach has the advantage of immediate feedback from the electrochemical stimulus and resulting chemistry. A combinatorial analysis of a gradient of driving forces is, thus, possible, due to the huge number of individual electrochemical cells. The combination of available X-Ray spectroscopic methods allows for unambiguous quantification yielding a non-destructive, non-contact *simultaneous* probe of driving force (actual reaction potential by XPS) and effect (electrochemistry by XAS) with a potential high throughput screening.

5. EC-PEEM example: the copper electroplating case

To demonstrate the operation of the electrochemical PEEM setup with the MCA platform, we have chosen aqueous copper (II) sulfate solution as a model system. This electrolyte is simple, well-behaved, well-studied, reversible, and allows the studying of the metal electroplating process. Additionally, it is more stable under the electron beam than electrolytes containing noble metal ions. All three elements found in $CuSO_4$, namely copper, oxygen, and sulfur, have pronounced peaks in the XAS energy range of interest and, therefore, their electrochemical transformations can be easily detected by PEEM. For these experiments, the MCA platform was filled with a 0.5 mol/L solution of $CuSO_4$ in 0.05 mol/L $H_2SO_4$. Due to the uncertainties in the filling/drying procedure, the electrolyte concentration in the microchannels was controlled within 30 % of its nominal value. As has been mentioned above, the graphene membrane and underlying gold coating on the front side of the MCA served as the working electrode and the platinum coating on the back of the channels formed a pseudo-reference electrode (RE). As the cell was cycled several times before/during PEEM measurements, metallic copper must have been deposited on the Pt RE, thus providing a reversible reference reaction. The cell was cycled between -1.25 V and 1.25 V vs. the RE to avoid extensive water electrolysis and metal deposition. The PEEM data were collected alongside cyclic voltammograms, at a fixed excitation energy (corresponding to either Cu L$_3$-edge or O K-edge) to yield three-dimensional datasets: partial electron yield intensity vs. lateral imaging position. BLU was used to process and visualize the data.

The photoelectron yield (PEY) intensity vs. photon energy spectra for copper and oxygen averaged over 100 individual microchannels are shown in Figures 5a,h. Figure 5b-g presents BLU results for a temporal PEEM dataset (PEY intensity vs. time) recorded at the photon energy of 931 eV (corresponding to the Cu$^{2+}$ ion L$_3$-edge adsorption peak; for peak assignment, see Ref.[41]). For simplicity, the dataset was unmixed into two components (on how to properly select the number of components, see Refs.[37, 42]): one potential-independent (Fig. 5b,d,f), and the other electrochemically-active (Fig. 5c,e,g). The inert component abundance map highlights the microscope aperture (circular yellow feature in Fig. 5b), where the PEEM signal is very low and constant. The central part of the image, showing both empty and electrolyte-filled microchannels, is blue, as it contains electrolyte and therefore generates a signal of a much higher intensity. The spectrum corresponding to this inactive background component is displayed in Figs. 5d,f alongside the used potential waveform and cyclic voltammogram (CV) recorded for the whole device. As can be seen, this component has low intensity and does not vary with the time or potential. The other component, however, (Fig., 5e,g) is strongly affected by the positive potential. As the WE potential is swept beyond ca. 0.5 V, the PEY intensity rapidly increases, until the potential is reversed, after which it slowly decreases to nearly initial value. The CV recorded for the whole device (Fig. 5g) has a clear anodic peak at 0.4 V and a broad, poorly-defined cathodic peak. After repeated cycling, the peaks slightly shift and increase in intensity (Fig. 5i, and Fig. 3c,f in Ref.[41]), indicating a stabilization of the reversible copper plating-stripping reaction proceeding at both electrodes. It should be noted that no exact match between the whole device CV and processes taking place in the individual channels should be expected. Variability between the channels' behavior is clearly seen in Fig. 5c. Thus, although the correlation between the onset of the rise in the PEY intensity in Fig. 5g and the anodic peak is not perfect, we still attribute both to the same process of metallic copper oxidation in two Marcus steps:

$$Cu^0 - e^- \rightarrow Cu^+ \text{ (slow)} \quad (1)$$

$$Cu^+ - e^- \rightarrow Cu^{2+} \text{ (fast)} \quad (2)$$

The increase of the PEY intensity is, then, due to the increase of the Cu$^{2+}$ ions concentration in the near-graphene region of the electrolyte due to reaction 2. Reduction of the bivalent copper ions during the cathodic process leads to the depletion of the PEEM-probed electrolyte layer of Cu$^{2+}$ and a subsequent decrease in the PEY (Fig. 5g). Interestingly, a similar behavior is observed for the data recorded at the photon energy of 541 eV, corresponding to the oxygen K-edge. The 3D dataset is again split into two components: the background (Fig. 5i,k,m) and electrochemically-active one (Fig. 5j,l,n). The strong correlation of the oxygen signal PEY with the anodic process is expected, since oxidation in an aqueous solution must increase oxygen atom density in the near-WE region. However, reaction 2 does not involve any oxygen-containing species, and thus cannot be directly responsible for this correlation. We attribute the increase in the oxygen signal during the anodic sweep to the adsorption of oxygen-carrying HSO$_4^-$ and SO$_4^{2-}$ ions on the graphene surface. This process must accompany reaction 2 and can explain the observed behavior. An alternative explanation could be the formation of copper (I) oxide, as proposed by Velasco-Velez *et al* in their recent publication:[43] $2Cu^+ + H_2O \rightleftarrows Cu_2O \downarrow + 2H^+$. Note, however, that this is not an electrochemical process, but simply hydrolysis, and at low pH the equilibrium of this reaction must be shifted to the left. The local equilibrium can also be affected by radiolysis products (as discussed below), especially, H$_2$O$_2$, favoring formation of copperoxide. However, we have not observed deposition of solids on the graphene membrane in this and similar experiments, thus, it is not likely that a Cu$_2$O precipitate is responsible for the observed oxygen signal increase.

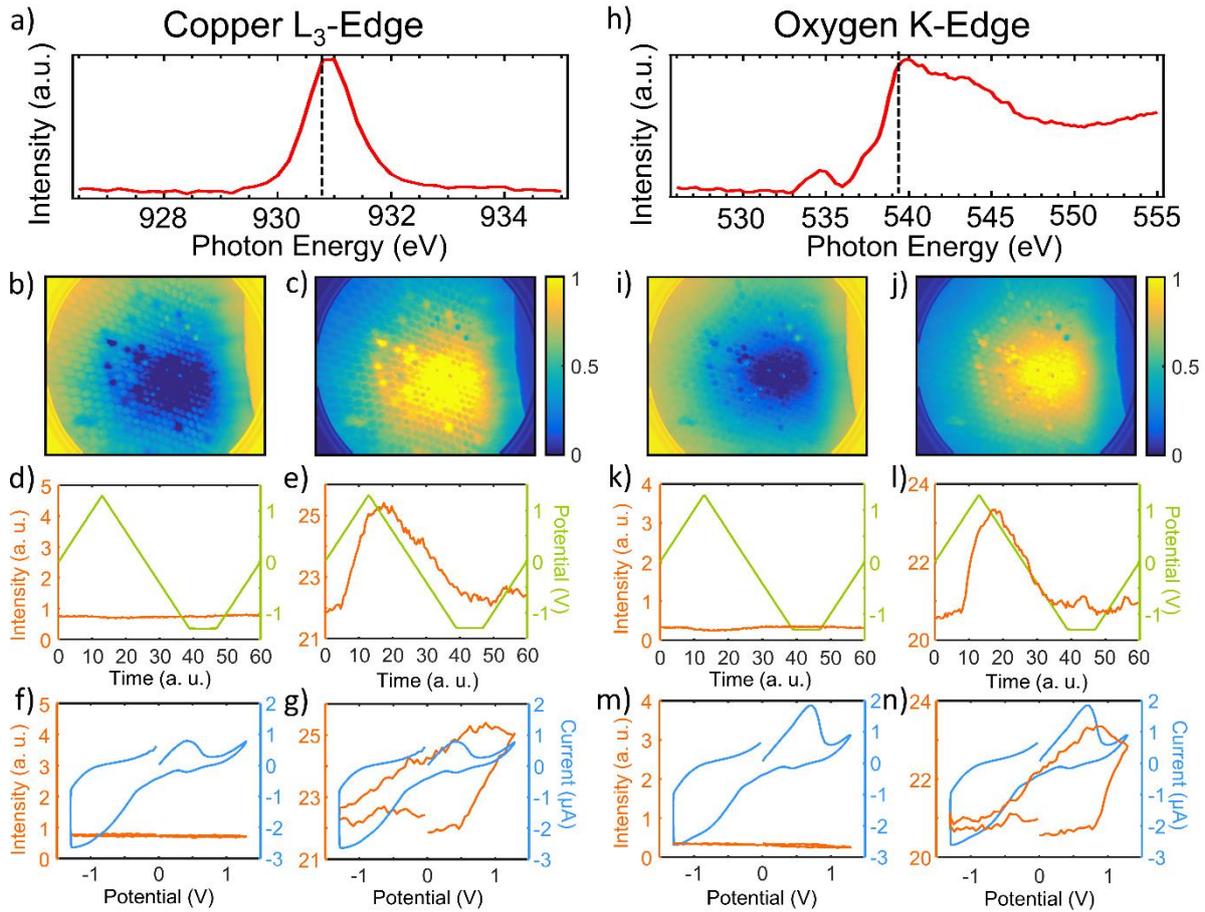

**Figure 5.** BLU of electrochemical PEEM datasets. Copper $L_3$ edge data (panels a-g): a) raw energy spectrum averaged over 100 microchannels, b) & c) extracted abundance maps of the background and electrochemically active components, respectively; the abundance maps show the spatial distribution of a particular component as a fraction of unity (0 % to 100 %); d) and e) are the corresponding BLU endmember components (PEEM intensity vs. time) plotted alongside with the WE potential variation; f) and g) are the same endmembers plotted vs. voltage and CVs recorded for the whole sample; Oxygen K-edge data (panels h-n): h) raw energy spectrum averaged over 100 microchannels, i) & j) extracted abundance maps of the background and electrochemically active components, respectively; k) and l) are the corresponding BLU endmember components (PEEM intensity vs. time) plotted alongside with the WE potential variation; m) and n) are the same endmembers plotted vs. voltage and CV's recorded for the whole sample. The spectral endmembers are plotted on the same scale of 4 units in the y-axis for ease of comparison.

The presented simple example of the PEEM electrochemical probing demonstrates, in principle, the possibility of such studies that are both local (probing electrolyte a few tens of angstroms below graphene) and spatially-resolved. As discussed above, the collection of statistics on the channels behavior should allow for a deeper understanding of the studied process, possible only due to high spatial resolution of the technique. To better demonstrate this point, the O K-edge dataset of Figures 5i-n was unmixed into 4 components (maximal meaningful number of the present behaviors). The loading maps in Figure 6b-d show that not all the filled channels behave the same way. The empty channels are mostly classified together with the aperture into the background component (Fig. 6a,e). The filled channels demonstrate 3 types of behaviors: strong PEY signal with weak potential dependence (Fig. 6f), strong PEY signal with strong potential dependence (Fig. 6h), and a large PEY signal with a very strong potential dependence (Fig. 6g). Localization of these behaviors (i.e. in which channels they occur) is visible in the corresponding loading maps (Fig. 6b-d). Whatever be the reasons for the observed spectral differences (variations in electrolyte concentration, contact quality, etc.), these maps exemplify the high importance of spatially-resolved studies capturing a statistically-significant collection of system responses, rather than only one of them.

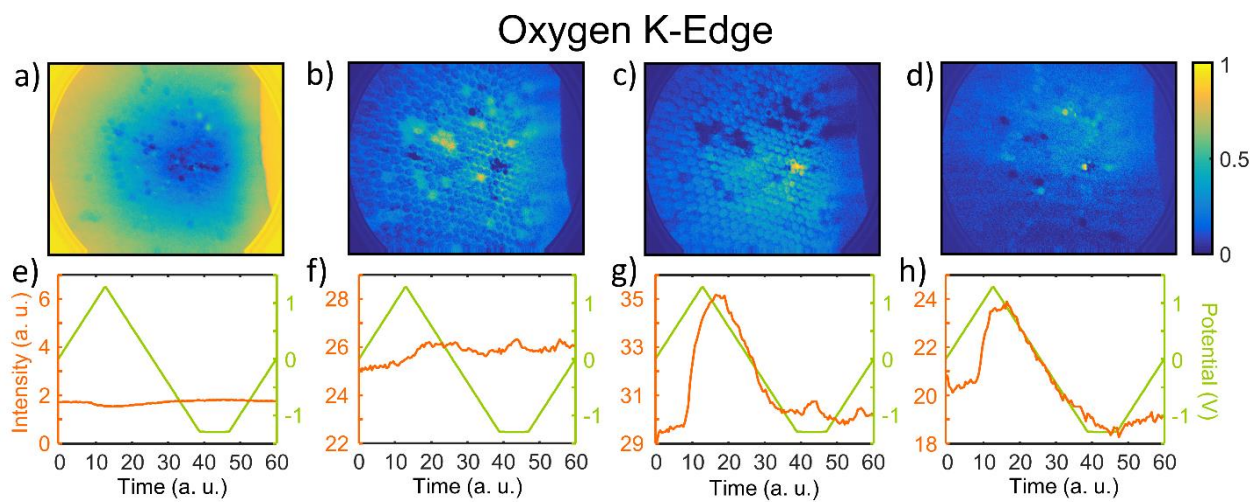

**Figure 6.** BLU of the Oxygen K-edge electrochemical PEEM datasets into 4 components: a)-d) loading maps; e)-f) corresponding spectral endmembers plotted alongside the WE potential. The spectral endmembers are plotted on the same scale of 7 units in the y-axis for ease of comparison.

The electrochemical PEEM measurements with the MCA platform are subject to a few challenges. First, the setup does not feature a stable reference electrode, with a controllable and reversible reaction on its surface. The current design of the MCA channels does not allow for accommodation of a standard reference electrode. The use of a Pt RE led to instabilities in the RE potential in the examples described above, as two competing processes took place on its surface: electrochemical splitting of water and copper plating and stripping. This results in both the broadness of the cathodic peaks and the shifts in the anodic peaks of the recorded CVs (Figs. 5c&i). Pre-deposition of a copper layer on the Pt RE would solve this problem for this particular system. The PEEM technique itself offers another solution to this problem: a unique capability of combining lateral and chemical sensitivity with electrochemical information. The XPS signal, which can be tuned to observe any chemical constituent, is a direct probe of the electric polarization[44, 45] manifesting itself in binding energy shifts. In the example shown in this work, the modulation of the difference of binding energies between the C1s (the working electrode) and the Cu2p (the reactant of the electrochemical reaction at the outer Helmholtz layer) directly reflects the reaction potentials of Eq. (1) and (2). Any ambiguity that might arise from the presence of more than one species ($Cu/Cu^+/Cu^{2+}$), is conveniently circumvented, as these are well distinguishable and quantifiable with simultaneous X-Ray absorption spectroscopy. For the general case of studying an electrochemical reaction proceeding at the graphene electrode and not involving copper or other metal plating, the reversible counter-reaction must be supplied at the Pt electrode. The counter-reaction with conversion potentials smaller than those of water splitting can be sustained by using a suitable concentration of an internal standard (i.e. ferrocene) in the electrolyte, thus providing a steady discharge rate at the Pt RE and suppressing its polarization that leads to potential shifts. The internal reference should: a) be stable in solution and form a good reversible pair, b) not interfere chemically or electrochemically with the main reaction under study (preferably, the internal standard should not contain same elements as in the substance under study), and c) be stable under irradiation.

The second challenge of the present MCA design is the slow evaporation of water via micro folds and wrinkles in the graphene membrane. The densely-packed structure of the micro-channels makes it possible for the adjacent channels to slowly exchange liquid through these wrinkles. Thus, a cell with ruptured membrane draws out electrolyte from the neighboring channels. The solution to this problem is to separate channels by larger distances, thus decreasing cross-talk between them. Indeed, the MCA design depicted in Figures 2c, g satisfies these conditions.

The third problem, inherent to any electron spectroscopy/microscopy technique, is radiation damage. In our experiments, we observed appreciable water radiolysis leading to the formation of hydrogen bubbles beneath the graphene membrane upon prolonged illumination. This led to a related slow decrease in the number of electrolyte-filled channels. The most abundant radiolysis products in the steady state are hydrogen and $H_2O_2$. These can unpredictably alter the local reaction conditions and electrolyte concentration, change the pH, and affect the graphene electrode surface. The oxidation of the graphene electrode or an increase in the defect density may strongly affect its electrochemical properties.[46] Hence, it is very important to define an artifact-free parameter space for irradiated electrolyte-graphene systems and limit the radiation dosage.

6. **Conclusions and outlook**

In summary, we have fabricated and successfully tested a new sample platform which enables imaging and spectroscopic probing of liquid-solid electrochemical interfaces using standard soft X-ray PEEM equipment. The advantage of this platform is its compatibility with UHV conditions while maintaining high transparency to the incoming X-rays and outgoing electrons due to an ultrathin molecularly impermeable graphene membrane. A multiplexed array-like nature of MCAs coupled with the wide field of view imaging and spectroscopic capabilities of PEEM has enabled the collection of statistically valuable spectroscopic and spatio-temporal data on the level of: (i) an ensemble of thousands of samples, (ii) each individual orifice within and (iii) each spatial pixel within the FOV. Using the Bayesian Linear Unmixing algorithm for dimensionality reduction, data denoising, and unmixing, we have analyzed the spectral and spatial information of liquid samples. We showed that high electric field of the PEEM has a negligible effect on interfacial ions distribution compared to standard electrochemical potentials. Using $CuSO_4$ aqueous solution as a model electrolyte, we demonstrated the power of PEEM to probe *in operando* the interfacial processes taking place correlatively with electrolyte polarization. Finally, considering recent results[47] we have discussed the limitations of the techniques associated with both beam damage effects and volume confinement. Several solutions to overcome the challenges have been proposed. We envision the expansion of this technique to modern ultrafast (laser, free electron laser excited) PEEM studies, as well as towards the coupling of the MCA platform to microfluidics and high pressure PEEM systems currently under development.

**Acknowledgements**

E.S., H.G., A.Y. acknowledge support under the Cooperative Research Agreement between the University of Maryland and the National Institute of Standards and Technology Center for Nanoscale Science and Technology, Award 70NANB14H209, through the University of Maryland. Heinz Pfeifer of Forschungszentrum Juelich and Jiri Libra of kolibrik.net were instrumental in the development of electrical devices and sample holders used in this publication. AT acknowledges CICECO-Aveiro Institute of Materials (Ref. FCT UID/CTM/50011/2013) financed by national funds through the FCT/MEC and, when applicable, co-financed by FEDER under the PT2020 Partnership Agreement.